\documentclass[12pt,a4paper]{iopart}
\usepackage{amsmath,amssymb}
\usepackage{adjustbox}
\usepackage[T1]{fontenc}
\usepackage[utf8]{inputenc}
\usepackage{textcomp} 
\usepackage{lmodern}
\usepackage{graphicx}
\usepackage{enumerate}
\usepackage{orcidlink}
\usepackage[dcu]{harvard}
\usepackage{longtable,booktabs,array}

\begin{document}

\title[Robust estimation of CO$_2$ airborne fraction]{Robust estimation of carbon dioxide airborne fraction under measurement errors}

\author{J. Eduardo Vera-Vald\'es$^*$, Charisios Grivas}

\address{Department of Mathematical Sciences, Aalborg University, Aalborg, DK}
\address{$^*$Corresponding author: eduardo@math.aau.dk}

\begin{abstract}

This paper discusses the effect of measurement errors in the estimation of the carbon dioxide (CO$_2$) airborne fraction. We are the first to present regression-based estimates and standard errors that are robust to measurement errors for the extended model, the preferred specification to estimate the CO$_2$ airborne fraction. To achieve this goal, we add to the literature in three ways: $i)$ We generalise the Deming regression to handle multiple variables. $ii)$ We introduce a bootstrap approach to construct confidence intervals for Deming regression in both univariate and multivariate scenarios. $iii)$ Propose to estimate the airborne fraction using instrumental variables (IV), taking advantage of the variation of additional measurements, to obtain consistent estimates that are robust to measurement errors. IV estimates for the airborne fraction are 44.8\%(± 1.4\%; 1$\sigma$) for the simple specification, and 47.3\%(± 1.1\%; 1$\sigma$) for the extended specification. We show that these estimates are not statistically different from the ordinary least squares (OLS) estimates, while being robust to measurement errors without relying on additional assumptions. In contrast, OLS estimates are shown to fall outside the confidence interval of the Deming regression estimates.

\end{abstract}
 

\vspace{-2pt}

\noindent{Supplementary material for this article is available \href{https://github.com/everval/Robust-CO2-Estimation-Supplementary}{online}}.



\maketitle

\section{CO$_2$ airborne fraction}\label{sec-intro}

The CO$_2$ airborne fraction is the portion of anthropogenic CO$_2$ emissions that remains in the atmosphere. This is an important factor in the carbon cycle, playing a critical role in assessing how human actions influence the climate system. Consequently, it attracts significant scientific interest. Previous studies on its estimation include \cite{canadell2007contributions,le2009trends,raupach2014declining,bennedsen2019trend,canadell2023intergovernmental,bennedsen2023evidence}. Recently, \cite{bennedsen2024} proposed a regression-based approach using ordinary least squares (OLS) to estimate the CO$_2$ airborne fraction. Their analysis demonstrates that, given mild assumptions, regression yields superior statistical properties compared to the conventional method of calculating it as a ratio. The authors report a point estimate of the airborne fraction over 1959–2022 of 47.4\%(± 1.1\%; 1$\sigma$) for their preferred specification including additional covariates.

A statistical challenge not fully explored by \cite{bennedsen2024} is the impact of measurement errors on OLS estimation. We show that measurement errors in emissions can bias the estimates of the CO$_2$ airborne fraction. To alleviate these concerns, \cite{bennedsen2024} used Deming regression \cite{deming1943} to obtain estimates robust to measurement errors. However, Deming regression demands some strong assumptions and presents several computational challenges.

On the one hand, Deming regression requires knowledge of the ratio of measurement error variances. In fields like chemistry, researchers might have a certain understanding of this ratio. This assumption is rarely met in the climate data. For example, \cite{bennedsen2024} considered five different values for this parameter. 

On the other hand, Deming regression assumes that the measurement errors follow a Gaussian distribution. This assumption is at odds with one of the main statistical properties of the OLS estimate suggested by \cite{bennedsen2024}. The authors motivated the use of OLS by the fact that the estimate follows a central limit theorem with a limiting Gaussian distribution, and the derivation does not require assuming that the error term has a Gaussian distribution. This property is not shared by the classical method based on a ratio, where Gaussianity must be assumed for the estimator to have a Gaussian distribution.

Moreover, Deming regression presents additional computational challenges. First, closed-form expressions for estimates in the multivariate setting are not available. This point is of great relevance given that the preferred specification of \cite{bennedsen2024} to estimate the CO$_2$ airborne fraction includes additional covariates in a multivariate specification. Second, standard errors and confidence intervals for Deming regression estimates are not available in closed-form expressions. The authors do not report on either for their estimates.

Inspired by these findings, this article contributes to the existing literature in multiple ways. First, it extends Deming regression to address the computational challenges. We introduce a straightforward approach to obtain estimates in the multivariate case and propose using bootstrap to compute standard errors and confidence intervals. Furthermore, we propose the use of instrumental variables (IV) \cite{reiersol1941confluence,durbin1954errors,sargan1958estimation} to estimate the CO$_2$ airborne fraction. IV is robust to measurement errors with a limiting Gaussian distribution without relying on a Gaussian assumption for the errors nor knowledge about their variances. In addition, IV can be easily extended to the multivariate setting and standard errors and confidence intervals can be obtained analytically. Hence, IV addresses all of the technical challenges of Deming regression and is our preferred method to estimate the CO$_2$ airborne fraction.

Our point estimate of the CO$_2$ airborne fraction is 44.8\%(± 1.4\%; 1$\sigma$) for the simple specification, and 47.3\%(± 1.1\%; 1$\sigma$) for the extended specification. We show that these estimates are not statistically different from the OLS estimates while being robust to measurement errors without relying on additional assumptions. Notably, given the discussion above on the Deming regression, this article is the first to provide estimates robust to measurement errors for the extended specification with additional covariates. 

\section{Data}\label{sec-data}

\begin{figure}[ht!]
    \centering
    \begin{tabular}{cc}
     \includegraphics[width=0.475\linewidth]{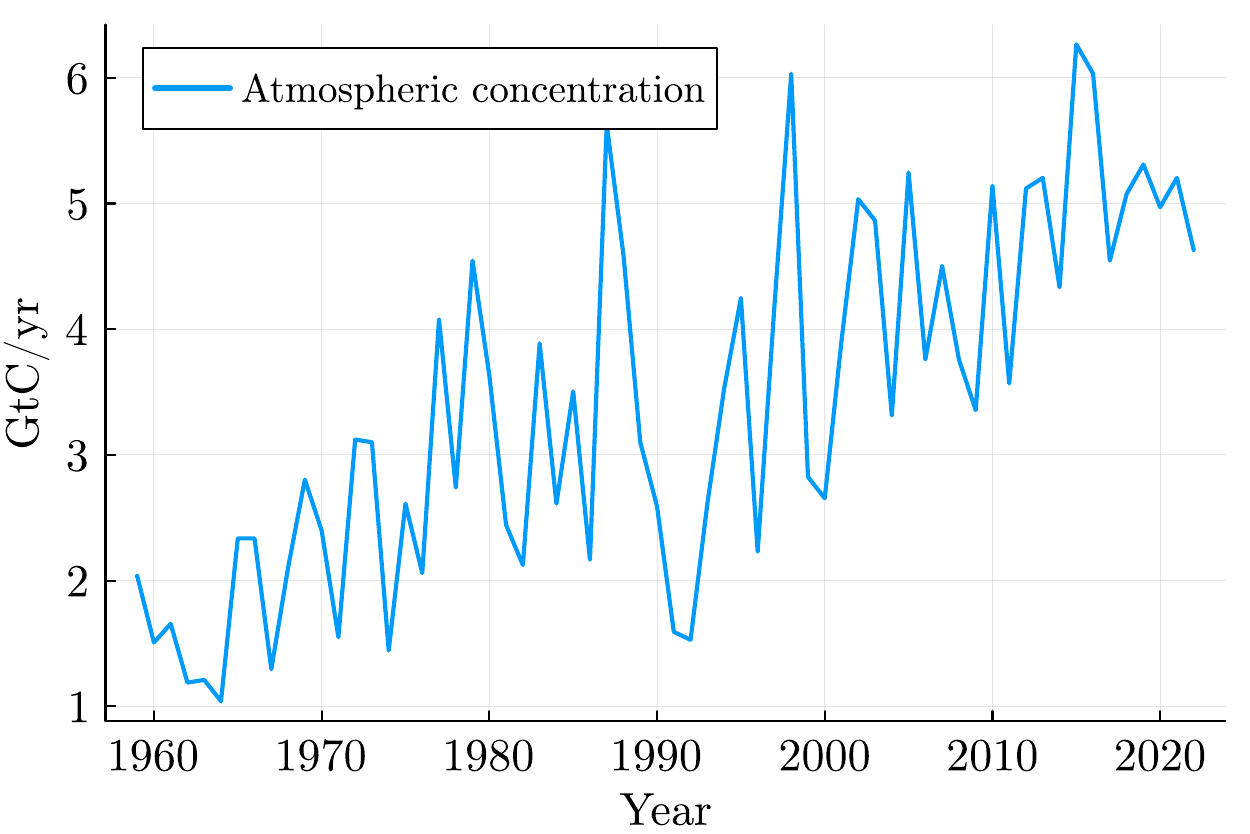} & \includegraphics[width=0.475\linewidth]{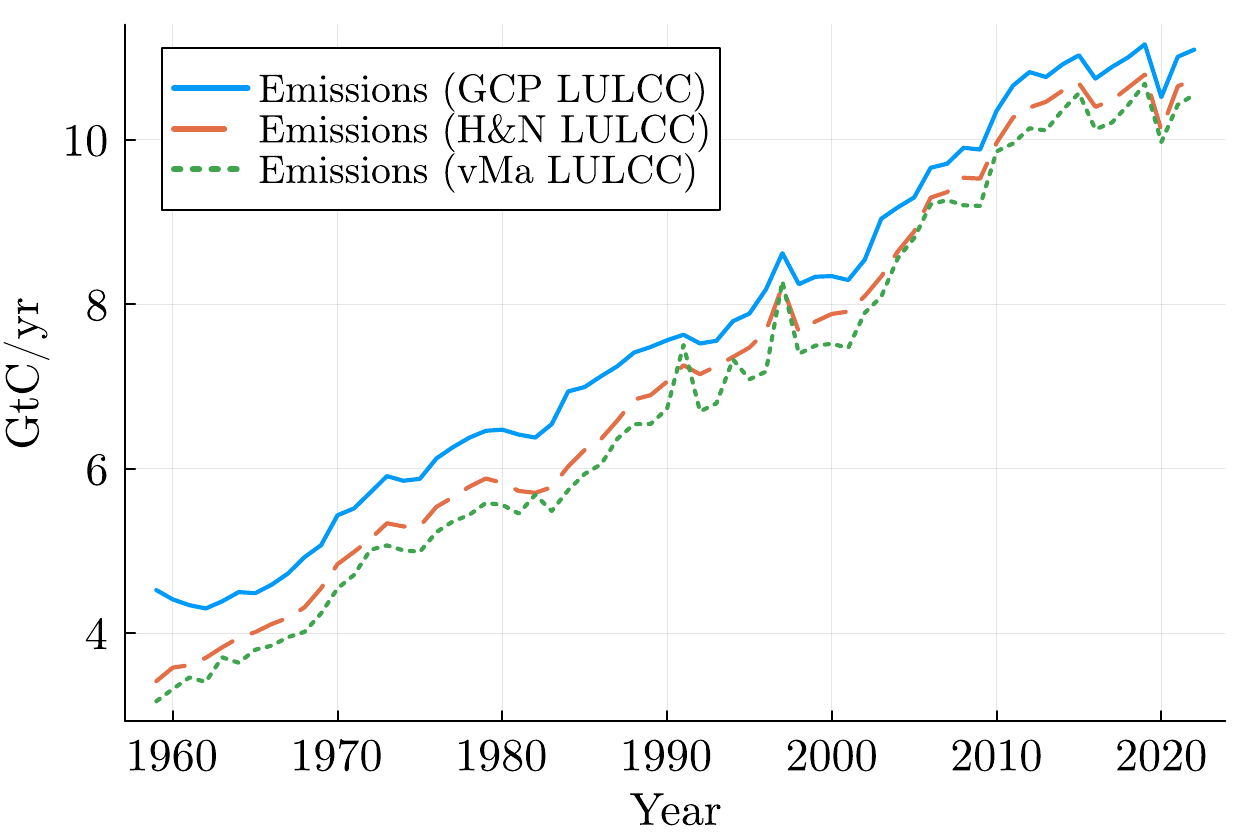} \\ 
     \includegraphics[width=0.475\linewidth]{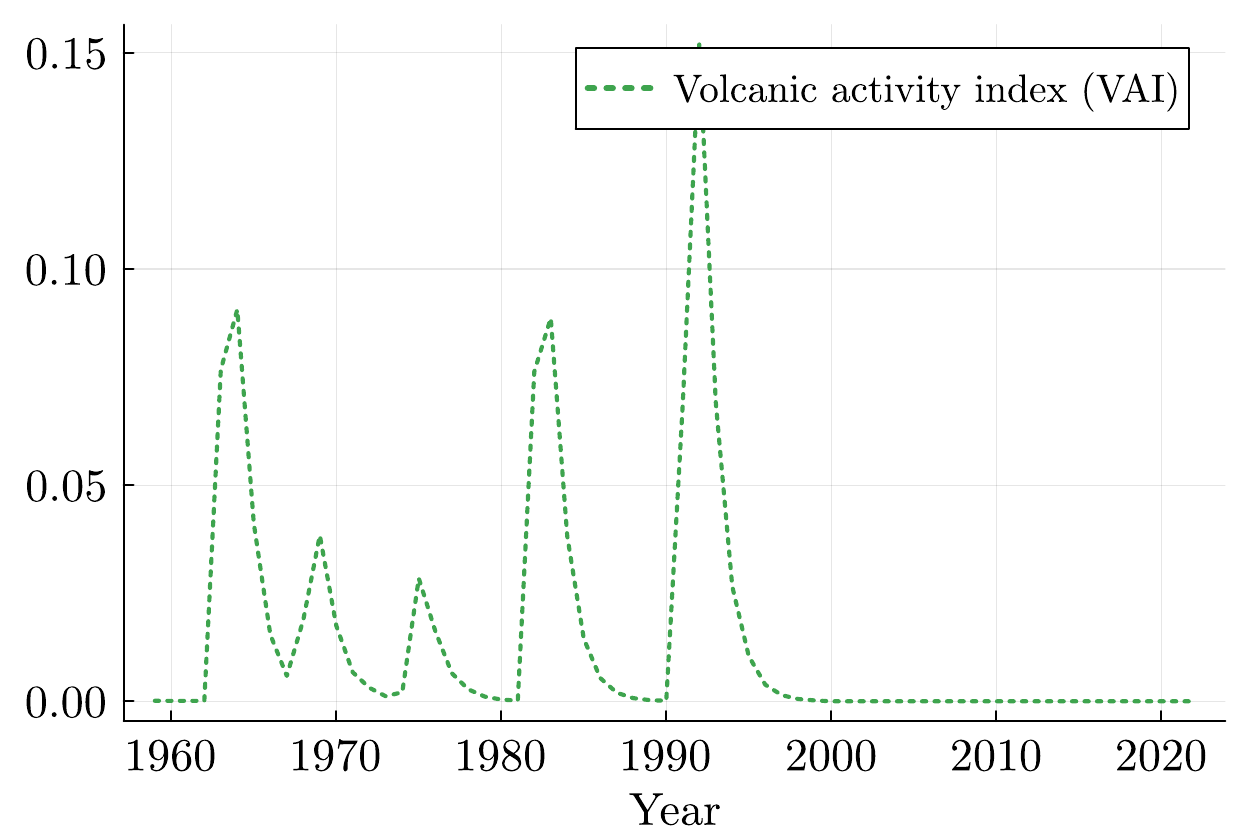} & \includegraphics[width=0.475\linewidth]{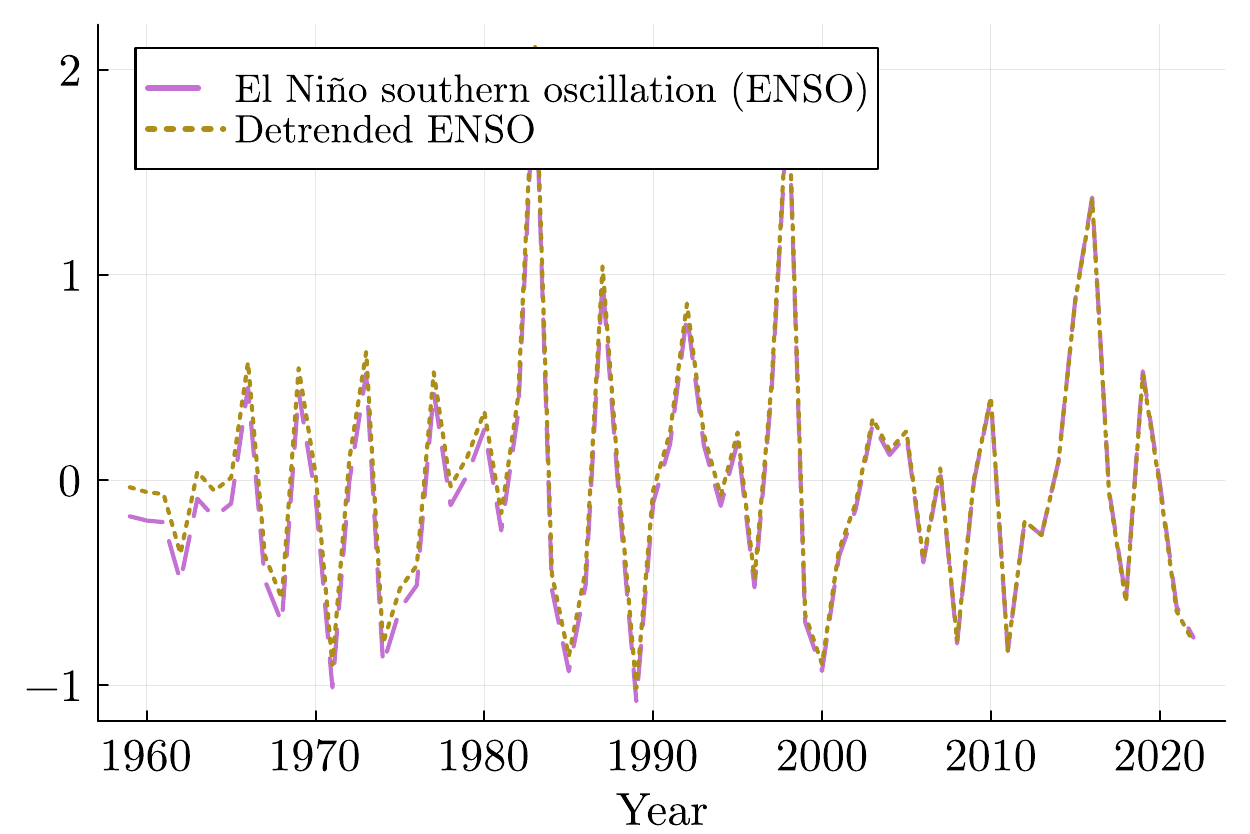} \\ 
    \end{tabular}
        \caption{Different land-use and land-cover change (LULCC) datasets [left]. Emission measures using different land use and land cover change (LULCC) datasets [right]}
    \label{fig-all-data}
\end{figure}

Figure~\ref{fig-all-data} shows the data used in this study. All data cover the period 1959–2022 and is measured yearly.

Data for atmospheric CO$_2$ [top left] and CO$_2$ emissions from anthropogenic sources [top right, blue solid] were obtained from the Global Carbon Project \cite{friedlingstein2023global}. Data for volcanic activity (VAI) [bottom left] are obtained from \cite{ammann2003monthly}. El Ni\~no Southern Oscillation (ENSO) data [bottom right, solid purple] are constructed by \cite{bennedsen2024} from the El Ni\~no 3 SST Index of the National Oceanic and Atmospheric Administration. 

Note that \cite{bennedsen2024} detrended the ENSO data [bottom right, dashed orange]. However, the trend is not statistically significant (p-value = 0.5464), so we use the original data for this study. Nevertheless, results using the detrended ENSO data are quite similar and can be recovered using the code in the supplementary material.

Additional data sources for emissions [top right, dashed orange and dotted green] are described in Section~\ref{Sec-measurement-errors}.

\section{Methods}\label{sec-methods}

\subsection{Linear models}

\cite{bennedsen2024} proposed to estimate the airborne fraction using OLS in the following linear model,
\begin{equation}\label{eq-simple-specification}
    G_{t}=\alpha E_{t}+u_{t}
\end{equation}
where $\alpha$ is the estimated CO$_2$ airborne fraction, $G_t$ are the changes in atmospheric CO$_2$, and $E_t$ are CO$_2$ anthropogenic emissions. The error $u_t$ is assumed to be a zero-mean-error process.

They also considered an extended specification including additional covariates to reduce the variance of the error process. The preferred specification of \cite{bennedsen2024} includes controls for the effects of the El Ni\~no Southern Oscillation and volcanic activity. Their extended specification is thus given by,
\begin{equation}\label{eq-extended-specification}
    G_{t}=\alpha E_{t} + \gamma_1 ENSO_t + \gamma_2 VAI_t +u_{t},
\end{equation}
where $ENSO_t$ is the El Ni\~no Southern Oscillation, $VAI_t$ is volcanic activity; and $G_t$, $E_t$, and $u_t$ are as before. 

For the rest of this paper, we will refer to equation~(\ref{eq-simple-specification}) and equation~(\ref{eq-extended-specification}) as the simple specification and extended specification, respectively. 

\subsection{Measurement errors}\label{Sec-measurement-errors}

The implications of measurement error can generally be divided into two cases based on severity. The first case involves measurement error only in the dependent variable ($G_t$ in equations (\ref{eq-simple-specification}) and (\ref{eq-extended-specification})). The second, which is of primary interest, involves the measurement error in the independent variable ($E_t$ in equations (\ref{eq-simple-specification}) and (\ref{eq-extended-specification})). As shown in the following, the measurement error in the dependent variable does not cause any bias in OLS, whereas the measurement error in the independent variable can have severe implications. 

We begin with the case where the dependent variable is the only variable measured with error. In equation~(\ref{eq-simple-specification}), assume that the changes in atmospheric CO$_2$ are measured with error; that is, $G_t=G^{\star}_t+\omega_t$ where $\omega_t$ is the measurement error. Substituting and rearranging terms, we obtain \begin{equation}\label{eq:IV1}
    G^{\star}_{t}=\alpha E_{t}+u_{t}-\omega_t
\end{equation}
which can be estimated by OLS provided that the population orthogonality condition $\mathbb{E}[E'(u-\omega)]=0$ holds. Note that OLS in equation~(\ref{eq:IV1}) will yield unbiased estimates since the model satisfies all the assumptions required. The only drawback is that the resulting residual will have a larger variance due to the presence of the additional term, compared to the estimate in equation~(\ref{eq-simple-specification}) where the true CO$_2$ coverage data are used. 

Next, we focus on the case of measurement error in the explanatory variable. One problem with estimating the airborne fraction is that the emissions data are potentially subject to measurement errors, particularly in the early years. Emissions are formally computed as $E_t = E_t^{FF}+E_t^{LULCC}$, where $E_t^{FF}$ are fossil fuel emissions, and $E_t^{LULCC}$ are the emissions from land-use and land-cover change (LULCC), respectively.

\begin{figure}[ht!]
    \centering
     \includegraphics[width=0.5\linewidth]{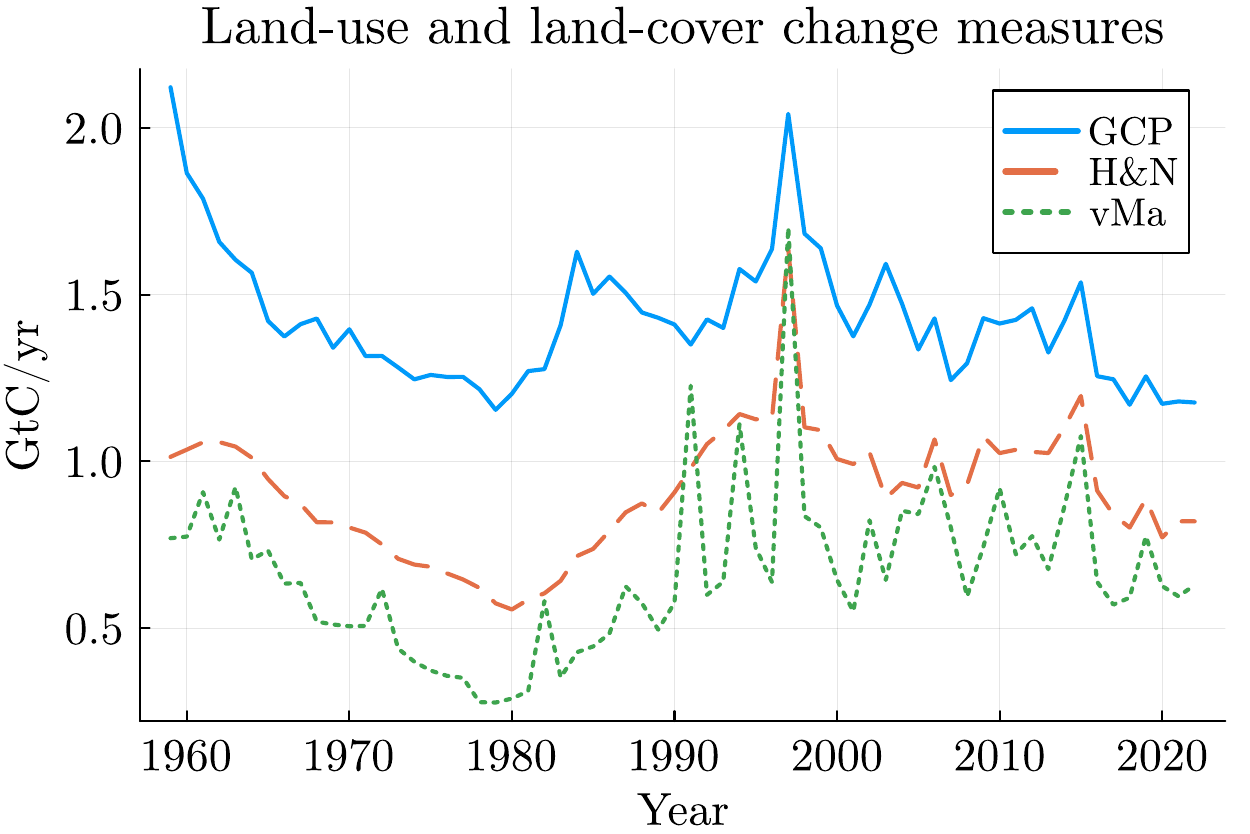} 
        \caption{Different land-use and land-cover change (LULCC) datasets.}
    \label{fig-lulcc-emissions}
\end{figure}

LULCC is measured using different methods. Figure~\ref{fig-lulcc-emissions} shows three different measurements for LULCC. The GCP LULCC data [solid blue] are from the Global Carbon Project \cite{friedlingstein2023global}, the H\&C LULCC data [dashed orange] are from \cite{houghton2022annual}, and the vMa LULCC data [dotted green] are from \cite{van2022retracted}. 

The measurement errors in LULCC translate into the measurement errors in the anthropogenic CO$_2$ emissions. Figure~\ref{fig-all-data} [top right] shows the different emissions measurements given the LULCC data used. Given the differences, it can be argued that there is some uncertainty in the measurements in the data. We show that measurement errors in emissions can bias the estimates of the CO$_2$ airborne fraction. 

Assume that we do not observe the true emissions, but rather a noisy version of it. That is, we observe $E_{1,t} = E_t^* + \eta_t$, where $E_t^*$ are the true emissions and $\eta_t$ is the measurement error, which we assume to have mean zero and variance $\sigma^2_\eta$. Estimating the CO$_2$ airborne fraction from noisy emissions data using OLS, we have:
\begin{equation}\phantomsection\label{eq-biased-estimator}{
\begin{aligned}
\hat{\alpha} & = (E_1^\top E_1)^{-1}(E_1^\top G) = \frac{\sum_{t=1}^T E_{1,t} G_t}{\sum_{t=1}^T E_{1,t}^2} = \frac{\sum_{t=1}^T (E_{1,t}^*G_t + \eta_t G_t)}{\sum_{t=1}^T (E_{1,t}^{*2}+2E_{1,t}^*\eta_t+\eta_t^2)}\\
            &\rightarrow \frac{\frac{1}{T}\sum_{t=1}^T E_{1,t}^*G_t}{\frac{1}{T}\sum_{t=1}^T E_{1,t}^{*2}+\sigma_\eta^2} = \alpha \times \left(\frac{\frac{1}{T}\sum_{t=1}^T E_{1,t}^{*2}}{\frac{1}{T}\sum_{t=1}^T E_{1,t}^{*2}+\sigma_\eta^2}\right),
\end{aligned}
}\end{equation}
where $\alpha$ is the true value for the CO$2$ airborne fraction. The notation on the second line denotes convergence as the sample size increases. Hence, the estimator does not converge to the true airborne fraction unless $\sigma_\nu^2=0$, which is tantamount to having no measurement errors. Furthermore, assuming that the measurement errors are not correlated with the variables of interest, equation~(\ref{eq-biased-estimator}) shows that the estimator is biased downward. The bias increases with the variance of the measurement error, which is unknown. 

\subsection{Deming regression}

To alleviate potential bias, \cite{bennedsen2024} used Deming regression \cite{deming1943}. As discussed in section~\ref{sec-intro}, Deming regression requires knowledge of the ratio of measurement error variances and the assumption that the errors are Gaussian. 

Formally, assume that $G_{t}$ and $E_{t}$ are noisy measurements of the variables $G^{\star}_{t}$ and $E^{\star}_{t}$, respectively, such that, 
\begin{align}
        G_{t}&=\alpha E_{t} + \epsilon_{G,t}, &\ \epsilon_{G,t}\overset{iid}{\sim} \mathbb{N}(0,\sigma^{2}_{G}), \label{eq-dem1} \\
	E_{t}&=E^{\star}_{t}+\epsilon_{E,t}, &\ \epsilon_{E,t}\overset{iid}{\sim} \mathbb{N}(0,\sigma^{2}_{E}) \label{eq-dem2}
\end{align}
where $iid$ stands for independent and identically distributed, $\mathbb{N}$ denotes the Gaussian distribution, and  $\epsilon_{G,t}$ and $\epsilon_{E,t}$ are assumed independent from one another. Let $\delta=\sigma^{2}_{G}/\sigma^{2}_{E}$ be the ratio of the two measurement error variances. The Deming regression estimate of $\alpha$ in the system defined by equations (\ref{eq-dem1})-(\ref{eq-dem2}) is obtained by maximum likelihood. It is given by, \begin{align}\label{eq:dem3} \hat{\alpha}_{Deming}=\frac{G^\top G-\delta E^\top E+\sqrt{(G^\top G-\delta E^\top E)^{2}+4\delta (E^\top G)^2}}{2E^\top G},
\end{align} 
which shows that the estimate directly depends on the chosen value for $\delta$. 
	
In addition to the theoretical hurdles, the Deming regression estimation presents some computational shortcomings. Estimates in the multivariate setting are not available in closed form. Furthermore, standard errors and confidence intervals for its estimates are not available analytically, even in the univariate specification. In the following, we develop methods to address these challenges.




\subsubsection{Deming regression multivariate case}\label{sec-multivariate-deming}

\hfill

To the best of our knowledge, no closed-form solution has been developed for the multivariate setting. Case in point, \cite{bennedsen2024} did not include results for the Deming regression for their preferred specification given by equation (\ref{eq-extended-specification}). Therefore, their analysis lacks results that are robust to measurement errors in the preferred specification. One aim of this paper is to address this shortcoming.

Consider the extended specification in equation~(\ref{eq-extended-specification}) where $G_t$ and $E_t$ are noisy measurements. We are interested in the system given by, 
\begin{align}
        G_{t}&=\alpha E_{t} + \gamma_1 ENSO_t + \gamma_2 VAI_t +\epsilon_{G,t}, &\ \epsilon_{G,t}\overset{iid}{\sim} \mathbb{N}(0,\sigma^{2}_{G}), \label{eq-mult-dem1} \\
	E_{t}&=E^{\star}_{t}+\epsilon_{E,t}, &\ \epsilon_{E,t}\overset{iid}{\sim} \mathbb{N}(0,\sigma^{2}_{E}), \label{eq-mult-dem2}
\end{align}
where all terms are as before.

In the supplementary material, we show that the Deming regression estimate for a multivariate system as in equations (\ref{eq-mult-dem1})-(\ref{eq-mult-dem2}) can be recovered from the univariate Deming regression using the Frisch-Waugh-Lovell (FWL) theorem \cite{frisch1933partial,lovell1963seasonal}.

In the context of the CO$_2$ airborne fraction, the FWL theorem guarantees that the CO$_2$ airborne fraction estimator in the preferred specification defined in equation~(\ref{eq-extended-specification}) is the same as the airborne fraction estimator in the following specification: \begin{equation}
    (\mathbb{I}-P_W)G_t = \alpha (\mathbb{I}-P_W)E_t + (\mathbb{I}-P_W)u_{t},
    \label{eq-extended-fwl}
\end{equation}
where $\mathbb{I}$ is the identity matrix of same size as the sample, $W_t = [ENSO_t,\ VAI_t]$ is the matrix containing the additional covariates, and $P_W = W(W^\top W)^{-1}W^\top$ is the projection matrix onto the column space defined by the additional covariates. That is, the FWL theorem shows that we can estimate the CO$_2$ airborne fraction using the residuals of regressing the atmospheric CO$_2$ concentration and CO$_2$ emissions on the El Ni\~no index and the volcanic activity index. 

Hence, estimating Deming regression in the multivariate system, equations (\ref{eq-mult-dem1})-(\ref{eq-mult-dem2}), is equivalent to using Deming regression in the system defined by equation~(\ref{eq-extended-fwl}) and equation~(\ref{eq-mult-dem2}). This novel theoretical result allows us to obtain Deming regression estimates robust to measurement errors for the preferred specification. 

Note, however, that the Deming regression assumptions of a Gaussian distribution for measurement errors and a known ratio of their variances are required.

\subsubsection{Deming regression inference}\label{sec:Bootstrap}

\hfill

Originally proposed by \cite{efron1992bootstrap}, bootstrap has become a major tool to approximate sampling distributions and variance of complex statistics. This is given its ability to estimate distributions for statistics when analytical solutions are unavailable. In addition, bootstrap methods often yield more accurate results than standard methods that rely on asymptotics. Hence, bootstrap is used as a means to improve the accuracy of confidence intervals.

For ease of exposition, we show how to employ a model-based bootstrap approach to calculate the confidence intervals of the Deming regression estimate, $\hat{\alpha}_{Deming}$, in the simple specification. The steps for the multivariate specification are analogous. The algorithm proceeds as follows:
\begin{enumerate}[i)]
	\item Estimate equation~(\ref{eq-simple-specification}) using the Deming regression formula shown in equation~(\ref{eq:dem3}) and obtain the residuals $\hat{u}_{t}$ for $t=1,\dots, T$ based on $\hat{\alpha}_{Deming}$. Let $\tilde{u}_{t}=\hat{u}_{t}-\frac{1}{T}\sum_{t=1}^{T}\hat{u}_{t}$ be the recentred residuals.
	\item Sample randomly with replacement the residuals $\tilde{u}_{t}$ to generate the bootstrap pseudo-residuals $\tilde{u}^{\ast}_{t}$. Generate pseudo-data in the $G$ domain by using recursively the following equation:\begin{equation}\label{eq:dem_boot}
		G^{\ast}_{t}=\hat{\alpha}_{Deming} E_{t}+\tilde{u}^{\ast}_{t}.
	\end{equation}
	\item Repeat the previous step B times, with B sufficiently large, and generate independent copies $\hat{\alpha}^{\ast}_{Deming,1},\dots,\hat{\alpha}^{\ast}_{Deming,B}$ based on equation~(\ref{eq:dem_boot}). 
        \item Calculate $$s.e(\hat{\alpha}_{Deming})=\sqrt{\frac{1}{B-1}\sum_{i=1}^{B}(\hat{\alpha}^{\ast}_{Deming,i}-\bar{\hat{\alpha}}^{\ast}_{Deming})^2},$$ where $\bar{\hat{\alpha}}^{\ast}_{Deming}=\frac{1}{B}\sum_{i=1}^{B}\hat{\alpha}^{\ast}_{Deming,i}$. The confidence intervals are then obtained as $$\left[\hat{\alpha}_{Deming}- q^{\ast}(1-\mathit{\alpha}/2)\ s.e(\hat{\alpha}_{Deming}),\hat{\alpha}_{Deming}+ q^{\ast}(\mathit{\alpha}/2)\ s.e(\hat{\alpha}_{Deming})\right],$$ 
	where $q^{\ast}(1-\mathit{\alpha}/2)$ and $q^{\ast}(\mathit{\alpha}/2)$  denote the $1-\mathit{\alpha}/2$ and $\mathit{\alpha}/2$ percentiles of $\hat{\alpha}^{\ast}_{Deming}$. 
\end{enumerate}	 

We use this bootstrap algorithm to obtain standard errors and confidence intervals for all Deming regression estimates considered in this study.


\subsection{Instrumental variable regression}

This article proposes estimating the airborne fraction using instrumental variables to obtain robust estimates without additional assumptions compared to those of OLS. That is, IV is a robust method that does not require the Gaussian assumption nor the knowledge of the measurement error variance. The trade-off is that IV requires the existence of an instrument that is correlated with the emissions variable but uncorrelated with the measurement error. Identifying instruments is a challenging task in most settings. However, in the context of CO$_2$ airborne fraction estimation, we propose using the different measurements of LULCC as emissions instruments. We show that we can use these different measurements to estimate the CO$_2$ airborne fraction without bias, even under the assumption that all of these different sources of data are subject to measurement error.

Consider an additional emissions measurement,
$E_{2,t} = E_t^* + \kappa_t$, where $\kappa_t$ is a measurement error, not correlated with $\eta_t$. Considering the different sources of LULCC data, it is reasonable to presume that the measurement errors do not correlate between them. Using $E_{2,t}$ as an instrument for $E_{1,t}$, IV estimates the airborne fraction as: \begin{equation}\phantomsection\label{eq-iv-estimator}{
\hat{\alpha}_{IV} = (E_2^\top E_1)^{-1}(E_2^\top G) = \frac{\sum_{t=1}^T E_{2,t}G_t}{\sum_{t=1}^T E_{2,t}E_{1,t}} = \frac{\sum_{t=1}^T (E_t^*G_t + \kappa_t G_t)}{\sum_{t=1}^T (E_t^{*2}+E_t^*(\eta_t+\kappa_t)+\eta_t\kappa_t)}\rightarrow\alpha.
}\end{equation}

Hence, $\hat{\alpha}_{IV}$ is not biased by the measurement error in the emissions variable. In addition to being unbiased, IV is consistent and follows a central limit theorem with limiting Gaussian distribution, even if the error terms are not Gaussian. Note that this last property is shared by OLS. In fact, the proof for the IV case closely follows that of the OLS case.


The instrumental variable estimator can be extended to the case where we have more than one instrument. In this case, we can use the generalised instrumental variable estimator (GIVE). Furthermore, the estimator can be easily applied to the extended model specification, which is the preferred specification in \cite{bennedsen2024}. Let $X$ be the matrix of regressors that includes emissions and additional covariates and let $Z$ denote the matrix of instruments. Then, the GIVE is computed as: \begin{equation}\label{eq-giv-estimator}
\hat{\alpha}_{GIVE} = (X^\top P_Z X)^{-1}(X^\top P_Z G)\rightarrow\hat{\alpha}^*,    
\end{equation}
where $P_Z = Z(Z^\top Z)^{-1}Z^\top$ is the projection matrix in the space defined by the instruments.

Similarly to IV, the GIVE is consistent and with limiting Gaussian distribution. The GIVE is also unbiased in the presence of measurement errors that are not correlated with the variables in the model. What is more, the variance of the GIVE has a closed-form expression given by: \begin{equation}\label{eq-giv-variance}
\mathrm{Var}\left[ {\hat{\alpha}_{GIVE}} \right] =  \hat{\sigma}^2 (X^\top P_Z X)^{-1},
\end{equation}
where $\hat{\sigma}^2 = \frac{1}{T} \sum_{t=1}^T (G_t - X_t\hat{\alpha}_{GIVE})^2$. 

For a textbook treatment of IV and GIVE, see \cite{davidson2004econometric}. Given all its statistical and computational advantages, GIVE is our preferred estimation method. To simplify notation, we denote both IV and GIVE as IV for the rest of this article.

\section{Results}

Table~\ref{tbl-panel-results-join} shows the estimates for the sample 1959-2022 (full sample) of the CO$_2$ airborne fraction using the LULCC measurements as instruments. Three cases are considered: $i)$ using H\&C LULCC as an instrument, $ii)$ using vMa LULCC as an instrument, and $iii)$ using both H\&C and vMa LULCC as instruments. The results for the simple specification are presented in the left columns, while the results for the extended specification are presented in the right most columns. As discussed in section~\ref{sec-data}, note that we used ENSO data without detrending, since there is no statistical evidence of a linear trend. However, the results using the detrended ENSO are quite similar. The table below also presents the results from the OLS and Deming regressions for both specifications.

\begin{table}[ht!]
\setlength{\tabcolsep}{8pt}
{\footnotesize
    \centering
    \caption{Ordinary least squares and instrumental variables estimates of the simple specification (left panel) and the extended specification including additional covariates (right panel) for the full sample (1959-2022). The 95\% confidence intervals for the OLS and IV regressions are based on the Gaussian distribution. The Deming regression estimates are obtained using $\delta \in \{0.2, 0.5, 1, 2, 5\}$. Standard errors and confidence intervals for Deming regression are computed using 9999 bootstrap replications.\\ }
    \begin{tabular}{l|lll|lll}
        \toprule
        Recent subsample &\multicolumn{3}{l|}{Simple specification, eqn. (\ref{eq-simple-specification})}&\multicolumn{3}{l}{Extended specification, eqn. (\ref{eq-extended-specification})}\\
        Model			&	Est.	&	S. e.	&	Conf.	interval	&	Est.	&	S. e.	&	Conf.	interval \\
        \midrule													
        OLS Regression			&	0.4478	&	0.0142	&	[0.4199,	0.4757]	&	0.4735	&	0.0108	&	[0.4522,	0.4947]	\\
        IV	reg.	(H\&N)	&	0.4479	&	0.0143	&	[0.4200,	0.4758]	&	0.4727	&	0.0108	&	[0.4514,	0.4939]	\\
        IV	reg.	(vMA)	&	0.4482	&	0.0143	&	[0.4202,	0.4761]	&	0.4723	&	0.0109	&	[0.4511,	0.4936]	\\
        IV	reg.	(H\&N-vMA)	&	0.4476	&	0.0142	&	[0.4197,	0.4756]	&	0.473	&	0.0109	&	[0.4515,	0.4944]	\\
        Deming	reg.	(0.2)	&	0.4623	&	0.0149	&	[0.4548,	0.4690]	&	0.4815	&	0.0107	&	[0.4760,	0.4865]	\\
        Deming	reg.	(0.5)	&	0.4561	&	0.0146	&	[0.4489,	0.4624]	&	0.4782	&	0.0107	&	[0.4728,	0.4831]	\\
        Deming	reg.	(1)	&	0.4526	&	0.0146	&	[0.4455,	0.4589]	&	0.4762	&	0.0106	&	[0.4698,	0.4811]	\\
        Deming	reg.	(2)	&	0.4504	&	0.0145	&	[0.4434,	0.4566]	&	0.4750	&	0.0105	&	[0.4698,	0.4798]	\\
        Deming	reg.	(5)	&	0.4489	&	0.0142	&	[0.4421,	0.4549]	&	0.4741	&	0.0107	&	[0.4688,	0.4789]	\\
        \bottomrule
    \end{tabular}
    \label{tbl-panel-results-join}
}
\end{table}

Several remarks are in order. First, estimates are centred around 44\% for the simple specification, while they are around 47\% for the extended specification. Second, the differences in estimates by OLS and IV are small. In the simple specification, the estimate using OLS is 44.78\%, while the IV estimate ranges from 44.76\% to 44.82\%. Additional tests in the supplementary material show that the estimates are not statistically different (p-values between 0.2672 and 0.8489). In contrast, the Deming regression, used by \cite{bennedsen2024} as a robust procedure for measurement error, varies from 45.04\% to 46.23\% depending on the selected measurement error variance ratio ($\delta$ in equation (\ref{eq:dem3})). 

Next, the right panel of table~\ref{tbl-panel-results-join} reports the results for the extended specification, equation (\ref{eq-extended-specification}), with ENSO and VAI included. Note that the results in \cite{bennedsen2024} do not consider this specification for the Deming regression. The OLS estimate is 47.35\%, while the IV estimates are 47.27\%, 47.23\%, and 47.30\%, depending on the instruments used. In contrast, the Deming estimates range from 47.41\% to 48.15\%. Hence, Deming regression estimates show more variability, due to the unknown measurement error variance ratio.

A significant contribution of our results above is that we can now quantify the uncertainty regarding the Deming regression estimates. It should be noted that the OLS estimates fall outside the confidence intervals of the Deming regression estimates using $\delta=0.2$ and $0.5$ for the simple specification, and for $\delta=0.2$ for the full specification. In contrast, the OLS estimate always falls within the confidence intervals of the IV estimates. These results show the sensitivity of Deming regression to the choice of $\delta$, and further illustrate our preference for IV estimation.

Table \ref{tbl-panel-results-join-recent} reports the results for the subsample starting in 1992, as a robustness check. In general, note that the smaller sample size increases the standard errors for all the estimates.

\begin{table}[ht!]
\setlength{\tabcolsep}{8pt}
{\footnotesize
    \caption{Ordinary least squares and instrumental variables estimates of the simple specification (left panel) and the extended specification including additional covariates (right panel) for the recent subsample (1992-2022). The 95\% confidence intervals for the OLS and IV regression are based on the Gaussian distribution. The Deming regression estimates are obtained using $\delta \in \{0.2, 0.5, 1, 2, 5\}$. Standard errors and confidence intervals for Deming regression are computed using 9999 bootstrap replications.\\}
    \centering
    \begin{tabular}{l|lll|lll}
        \toprule
        Recent subsample &\multicolumn{3}{l|}{Simple specification, eqn. (\ref{eq-simple-specification})}&\multicolumn{3}{l}{Extended specification, eqn. (\ref{eq-extended-specification})}\\
        Model			&	Est.	&	S. e.	&	Conf.	interval	&	Est.	&	S. e.	&	Conf.	interval \\
        \midrule												
        OLS Regression		&	0.4497	&	0.0173	&	[0.4157, 0.4836]		&	0.4613	&	0.0112	&	[0.4392, 0.4833]		\\
        IV	reg.	(H\&N)	&	0.4496	&	0.0173	&	[0.4157, 0.4836]		&	0.4622	&	0.0112	&	[0.4402, 0.4842]		\\
        IV	reg.	(vMA)	&	0.4502	&	0.0245	&	[0.4022, 0.4982]		&	0.4623	&	0.0112	&	[0.4402, 0.4843]		\\
        IV	reg.	(H\&N-vMA)	&	0.4495	&	0.0173	&	[0.4156, 0.4834]		&	0.4622	&	0.0112	&	[0.4401, 0.4842]		\\
        Deming	reg.	(0.2)	&	0.4598	&	0.0176	&	[0.4509, 0.4675]		&	0.4662	&	0.0105	&	[0.4611, 0.4709]		\\
        Deming	reg.	(0.5)	&	0.4555	&	0.0175	&	[0.4468, 0.4630]		&	0.4645	&	0.0106	&	[0.4594, 0.4692]		\\
        Deming	reg.	(1)	&	0.4531	&	0.0172	&	[0.4446, 0.4603]		&	0.4636	&	0.0107	&	[0.4584, 0.4683]		\\
        Deming	reg.	(2)	&	0.4515	&	0.0172	&	[0.4431, 0.4587]		&	0.4630	&	0.0107	&	[0.4578, 0.4677]		\\
        Deming	reg.	(5)	&	0.4504	&	0.0171	&	[0.4422, 0.4576]		&	0.4625	&	0.0106	&	[0.4574, 0.4672]		\\
        \bottomrule
    \end{tabular}
    \label{tbl-panel-results-join-recent}
}
\end{table}

In the simple specification, the findings align closely with those of the full sample: the OLS and IV estimates are similar. The Deming regression produces higher estimates that decrease as $\delta$ increases. As in the full sample, the OLS estimate falls outside the confidence interval of the Deming regression for $\delta=0.2$. 

The extended specification for the subsample yields lower estimates than the extended specification for the full sample. The OLS estimate is 46.13\%, while the IV estimates are 46.22\% and 46.23\%. Deming regression estimates range from 46.25\% to 46.62\%. In this case, in part due to the larger standard errors, the OLS estimate falls within the confidence interval of the Deming regression. 

\section{Discussion}

 The CO$_2$ airborne fraction estimates in the literature have fluctuated around a constant value over the period 1959 to 2022. The consensus estimate of the CO$_2$ airborne fraction is around 44.0\%. \cite{bennedsen2024}, using a regression-based approach in an extended specification, found this parameter to be around 47.0\%. In this article, we examine the impact of measurement errors on the estimation of the CO$_2$ airborne fraction. Anticipating possible bias given measurement errors, \cite{bennedsen2024} used Deming regression. However, Deming regression was only used in the simple specification and no standard errors or confidence intervals were computed. Furthermore, Deming regression relies on strong assumptions about the measurement errors.

 To alleviate these shortcomings, in this paper we develop a method to estimate the Deming regression in the extended specification. Furthermore, we show how to use bootstrap to obtain standard errors and confidence intervals. We are the first to obtain estimates, standard errors, and confidence intervals for the multivariate Deming regression specification. 
  
 Given the strong assumptions required for Deming regression, we propose using IV and show how to use different measurements as instruments. Our results show that IV provides estimates of the CO$_2$ airborne fraction that are close to the OLS estimates. However, IV remains unbiased in the presence of measurement errors. Moreover, in contrast to the Deming regression case, there are closed-form expressions to compute the standard errors for IV. Hence, our results add robustness to the estimation of the CO$_2$ airborne fraction in the presence of measurement errors.

\section*{Reproducibility statement}

All the code and data used in this analysis can be accessed in the corresponding author's GitHub repository at:\\ \url{https://github.com/everval/Robust-CO2-Estimation-Supplementary}. 

The authors believe that open source code is essential for inclusive research, particularly in the context of climate change, which is a global challenge that affects all countries, especially developing countries. Hence, all the results presented in this paper are obtained using Julia \cite{julia}, a free open source programming language. 

\section*{Acknowledgements}
The authors thank Mikkel Bennedsen, Eric Hillebrand, and Olivia Kvist for useful comments and suggestions to an earlier draft. 

\section*{Conflict of interest}
The authors declare that they have no competing interests.
\phantomsection\label{refs}

\section*{References}

\bibliography{references}

\end{document}